\documentclass{article}
\usepackage{ijcai23}

\usepackage{makecell}

\usepackage{graphicx}
\usepackage{paralist}
\usepackage{url}
\usepackage{xspace}
\usepackage{amsfonts,amsmath}
\usepackage{booktabs}
\usepackage{setspace}
\usepackage{multirow}
\usepackage{multicol}
\usepackage[hidelinks]{hyperref}
\usepackage{amssymb}%
\usepackage{pifont}%
\usepackage{color, colortbl}
\usepackage{bbm}
\usepackage{wrapfig}
\usepackage{lipsum}
\usepackage{enumitem}
\usepackage{paralist, tabularx}

\usepackage{times}

\usepackage{soul}
\usepackage{url}
\usepackage[utf8]{inputenc}
\usepackage[small]{caption}
\usepackage{graphicx}
\usepackage{amsmath}
\usepackage{booktabs}
\usepackage{stmaryrd}
\usepackage{times}

\usepackage{soul}
\usepackage{url}
\usepackage{graphicx}
\usepackage{amsmath}
\usepackage{stackrel,amssymb}
\usepackage{booktabs}
\usepackage{balance}
\usepackage{graphics}
\usepackage{graphicx}
\usepackage{multirow}
\usepackage{caption}
\usepackage{makecell}
\newcommand*{\scale}[2][4]{\scalebox{#1}{$#2$}}

\title{Graph-based Molecular Representation Learning}

\author{
Zhichun Guo$^1$, Kehan Guo$^1$, Bozhao Nan$^1$, Yijun Tian$^1$, Roshni G. Iyer$^2$, Yihong Ma$^1$, \\
Olaf Wiest$^1$, Xiangliang Zhang$^1$, Wei Wang$^2$, Chuxu Zhang$^3$, Nitesh V. Chawla$^1$\\
\affiliations
$^1$University of Notre Dame\\
$^2$University of California, Los Angeles\\
$^3$Brandeis University
\emails
\large{\{zguo5,kguo2,bnan,yijun.tian,yma5,Olaf.G.Wiest.1,xzhang33,nchawla\}@nd.edu\\\{roshnigiyer,weiwang\}@cs.ucla.edu, chuxuzhang@brandeis.edu}
}

\begin{document}

\maketitle

\begin{abstract}
Molecular representation learning (MRL) is a key step to build the connection between machine learning and chemical science. In particular, it encodes molecules as numerical vectors preserving the molecular structures and features, on top of which the downstream tasks (e.g., property prediction) can be performed. Recently, MRL has achieved considerable progress, especially in methods based on deep molecular graph learning. In this survey, we systematically review these graph-based molecular representation techniques, especially the methods incorporating chemical domain knowledge. Specifically, we first introduce the features of 2D and 3D molecular graphs. Then we summarize and categorize MRL methods into three groups based on their input. Furthermore, we discuss some typical chemical applications supported by MRL. To facilitate studies in this fast-developing area, we also list the benchmarks and commonly used datasets in the paper. Finally, we share our thoughts on future research directions. 


\end{abstract}

\vspace{-0.14in}
\section{Introduction}

The interaction between machine learning and chemical science has received great attention from researchers in both areas. Remarkable progress  has been made by applying machine learning in various chemical applications including molecular property prediction~\cite{guo2020graseq,sun2021mocl,yang2021deep,liu2022spherical}, reaction prediction~\cite{jin2017predicting,do2019graph}, molecular graph generation~\cite{jin2018junction,jin2020hierarchical} and drug-drug interaction prediction~\cite{lin2020kgnn}. Molecular representation learning (MRL) is an important step in bridging the gap between these two fields. MRL aims to utilize deep learning models to encode the input molecules as numerical vectors, which preserve relevant information about the molecules and serve as feature vectors for downstream chemical applications. While general representation learning models were earlier adapted to represent molecules, MRL algorithms have been recently designed to better incorporate chemical domain knowledge. However, it is non-trivial to have a seamless integration of domain knowledge into representation learning models. Given the tremendous effort in this rapidly-developing area, we are motivated to provide a systematic review of recent MRL methods, which are based on graph deep learning methods and integrate various types of chemical domain knowledge.

We focus on graph-based MRL for two reasons. First, molecules naturally lead themselves to graph representations, as they are essentially atoms and bonds interconnecting atoms. Compared with SMILES, a line-based representation (i.e., string) of molecules, molecular graphs provide richer information for MRL models to learn from. Accordingly, graph-based MRL models evolve much faster than sequence-based MRL models. Second, graph neural networks (GNN) have shown exceptional capacity and promising performance in handling graph structural data~\cite{kipf2016semi,hamilton2017inductive,zhang2019heterogeneous,guo2022boosting}, certainly including those applied on molecular graphs~\cite{gilmer2017neural,Hu2020Strategies,you2020graph}. It is thus urgent to summarize the effort of leveraging GNN on molecular graphs with domain knowledge and open this topic with more discussion.

This survey paper will be a contribution to both fields of machine learning and chemistry. A variety of molecule-centered problems can be formulated as predictive or generative tasks, e.g., molecule property prediction, reaction prediction, and molecule generation. The machine learning-enabled solutions for these problems have a common ground in learning high-quality representations for molecules. However, researchers in chemistry are overwhelmed by the big group of MRL models to choose from, not to mention that new models are rapidly presented. This survey provides an up-to-date overview of MRL models regarding the input graph to the models and the feasible downstream applications. Researchers in chemistry can easily find out the MRL models that match their application needs. For researchers in machine learning, a lack of understanding of the chemical domain knowledge is the barrier to addressing the representation learning for molecules. Treating molecular graphs as regular attributed graphs would overlook the special substructure patterns of molecules, such as motifs and functional groups. This survey summarizes the existing strategies for introducing chemistry-related domain knowledge into representation learning and will inspire researchers in machine learning to design more effective MRL models.

We organize the review in the following structure. We first introduce the input expression of molecules by 2D and 3D graphs. Following these categories of input, we summarize representative MRL algorithms regarding the usage of domain knowledge, the learning strategies, the application tasks, and code links if available. We then look into each application task and introduce in detail the usable MRL models and benchmark datasets. In the end, we open the discussion for future research directions and conclude.

\vspace{-0.15in}
\section{Expression of Molecules }
\begin{figure}[t]
    \centering
    {\includegraphics[width=0.46\textwidth]{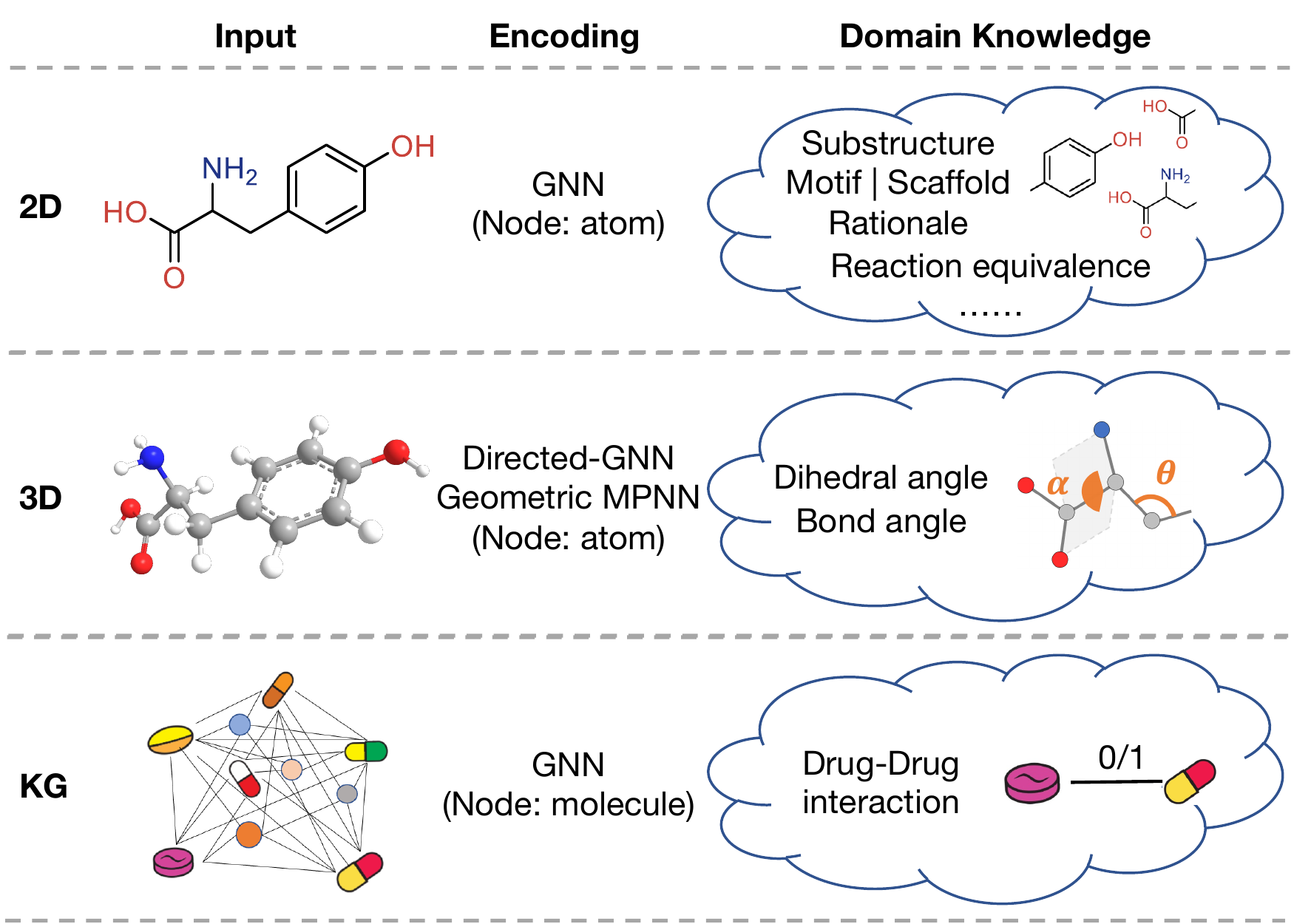}}
    \caption{Overview of graph-based MRL.}
\label{fig:view}
\end{figure}

When machine learning was first introduced for molecular analysis, hand-crafted features based on predetermined fingerprint extraction rules were used to identify and represent significant information about molecules \cite{ahneman2018predicting}.

However, this feature engineering process can be time-consuming, expert-dependent, and may not always provide the best results. To address these challenges, deep learning models, known for representation learning, have been developed to learn important molecular features automatically. Molecule input to deep learning models can be presented in two kinds of expressions: molecular sequences and graphs. 

The expression of molecular sequence, such as simplified molecular-input line-entry system (SMILES)~\cite{weininger1989smiles} and SELF-referencing Embedded Strings (SELFIES)~\cite{krenn2020self}, may separate two connected atoms at two distant positions and lead to inferior representation. 

In contrast, the expression by graph naturally incorporates additional information in nodes (atoms) and edges (bonds), which can be easily leveraged by the rich suite of graph-based models (e.g., graph neural networks). Therefore, MLR on molecular graphs is becoming commonly used and will be the focus of this survey.

In this section, we provide clarification on the distinction between 2D molecular graphs and 3D molecular graph representations, as shown in Figure~\ref{fig:view}. We analyze the characteristics of each representation and explore their applications and limitations when used in deep learning models.

\subsection{2D Molecular Graphs}
A graph is typically composed of nodes connected by edges. Similarly, in a molecule, atoms can be seen as nodes and bonds as edges interconnecting them. Thus, each molecule has a natural graph structure. This renders molecular graphs to be the most feasible input for deep learning models and leads to their extensive use. The most common form of molecular graphs is described by three matrices: the node feature matrix, edge feature matrix, and adjacency matrix. Molecules are usually stored as SMILES for convenience and then converted to molecular graphs for computation using specific tools, such as RDKit~\cite{rdkit}. The commonly used features of nodes and edges are listed in Table~\ref{tab:graph_data}, including mandatory features such as atom and bond types, and other optional features can be added as in need of different tasks~\cite{tang2020self,saebi2023use}. Among these features, the atom's chirality tag cannot be learned from the common 2D molecular graph representation without 3D geometric information, while all other features are learnable from both 2D and 3D structures. Each bond is considered as a bidirectional edge, i.e., a bond between atoms A and B is given as two edges in the adjacency list: one from A to B, and the other from B to A. With the description matrices, 2D molecular graphs can be treated as homogeneous~\cite{gilmer2017neural,guo2021few,coley2019graph} or heterogeneous networks~\cite{shui2020heterogeneous} for learning molecular representations via leveraging graph neural networks (GNN).

\begin{table}
\begin{center}
\scale[0.8]{{\begin{tabular}{lll}
\toprule
\textbf{Attribute} &  \textbf{Details}\\ 
\midrule
\textbf{Node} \\
\ \ Atom type & 118\\
\ \ Chirality tag & unspecified, tetrahedral cw, tetrahedral ccw, other\\
\ \ Hybridization & sp, sp$^2$, sp$^3$, sp$^3d$, or sp$^3$d$^2$\\
\ \ Atomaticity & 0 or 1 (aromatic atom or not)\\
\textbf{Edge}\\
\ \ Bond type & single, double, triple, aromatic\\
\ \ Ring & 0 or 1 (bond is in a ring or not)\\
\ \ Bond direction & -, endupright, enddownright\\
\ \ Stereochemistry & -, any, Z, E, cis, trans, @\\
\bottomrule
\end{tabular}}}
\caption{Details of node and edge features in molecular graphs.}
\label{tab:graph_data}
\end{center}
\end{table}

Despite that GNN can be conveniently used on 2D molecular graphs, the learned representation neglects the spatial direction and torsion between atoms in molecules. This is mainly due to the limitation of 2D graph structure, which only presents the type of bond connecting two atoms, and has no information like torsion angle, bond length, and stereoisomerism. The missing information is important to catch the subtle difference that may cause a significant discrepancy in chemical problems. For instance, reactants with the same 2D graph structures can have different products, because the reactants may differ on torsion angle and bond length, which are not reflected in their 2D graphs.  

\subsection{3D Molecular Graphs}
The 3D molecular graphs provide the missing geometric information by explicitly encoding the spatial structure of molecules. The 3D graphs present the atomic structure as a set of atoms along with their 3D coordinates, which includes more spatial information about atoms. As a result, this representation format has received increasing attention in MRL~\cite{liu2022spherical}. The key difference between 2D and 3D molecular graphs lies in the determination of edges between atoms. In 2D molecular graphs, edges (bonds) are pre-determined, indicating the existence of a connection or not. The soft edges of atomic interactions in 3D graphs can be determined by the distance and also the angle between two atoms using their coordinates.

To incorporate more complicated spatial relationships, spherical graph neural network~\cite{liu2022spherical} is designed to learn molecule structure from 3D graphs.

\begin{table*}[t]
\begin{center}
\scale[0.76]{{\begin{tabular}{c|l|l|l|l|l|l|l}
\toprule
\textbf{Input} & \textbf{Algorithm} & \textbf{Encoder} &\textbf{Pretrain} & \textbf{Domain Knowledge} 
& \textbf{Tasks} 
&\textbf{Venue} &\textbf{Code Link} \\ 
\midrule
\multirow{19}{*}{2D} &MPNN$^{[1]}$ &MPNN & / & / &PP& ICML'17 & / \\
 &Pre-GNN$^{[2]}$ &GIN & RT &/ & PP&ICLR'20 &\url{https://github.com/snap-stanford/pretrain-gnns/}\\
 &InfoGraph$^{[3]}$ &GNN &CL&/ & PP&ICLR'20   &\url{https://github.com/fanyun-sun/InfoGraph}\\
&GNN-FiLM$^{[4]}$ & MPNN &/ &/&PP & ICML'20  &\url{https://github.com/microsoft/tf-gnn-samples}\\
 &GROVER$^{[5]}$ &GAT &  RT &Motif&PP&NeurIPS'20 &\url{https://github.com/tencent-ailab/grover}\\
 &GraphCL$^{[6]}$ &GNN &CL  &/ &PP&NeurIPS'20  &\url{https://github.com/Shen-Lab/GraphCL}\\
 &MoCL$^{[7]}$ &GIN &CL   &Motif\&General carbon&PP & KDD'21  &\url{https://github.com/illidanlab/MoCL-DK}\\
 &MGSSL$^{[8]}$ &GIN &  RT &Motif&PP & NeurIPS'21  &\url{https://github.com/zaixizhang/MGSSL}\\
 &PhysChem$^{[9]}$ &MPNN &/ & PhyNet and ChemNet &PP& NeurIPS'21  & /\\
&GERA$^{[10]}$ &GNN & / &Rationale &PP&KDD'22&\url{https://github.com/liugangcode/GREA}\\
&MoleOOD$^{[11]}$ &SAGE &/ &Scaffold & PP&NeurIPS'22&\url{https://github.com/yangnianzu0515/MoleOOD}\\
&JT-VAE$^{[12]}$ &MPNN & /&Chemical substructure &MG& ICLR'18  &\url{https://github.com/wengong-jin/icml18-jtnn} \\
&VJTNN$^{[13]}$ &MPNN  & /&Chemical substructure &MG&ICLR'18  &\url{https://github.com/wengong-jin/iclr19-graph2graph}\\
&GraphAF$^{[14]}$ &R-GCN  & /&Valency constraint &MG& ICLR'20  &\url{https://github.com/DeepGraphLearning/GraphAF}\\
&RationaleRL$^{[15]}$ &MPNN & /&Rationale & MG&ICML'20  &\url{https://github.com/wengong-jin/multiobj-rationale}\\
&HierVAE$^{[16]}$ &MPN  &/ &Motif &MG& ICML'20  &\url{https://github.com/wengong-jin/hgraph2graph}\\
&MoLeR$^{[17]}$ &GNN &  /&Motif &MG&ICLR'22 & /\\
&WLDN$^{[18]}$ &WLN & / &/ & RP&NeurIPS'17  &\url{https://github.com/wengong-jin/nips17-rexgen} \\
&MolR$^{[19]}$ &GNN &CL&Reaction equivalence&RP & ICLR'22  & \url{ https://github.com/hwwang55/MolR}\\
\midrule
\multirow{7}{*}{3D} & DimeNet$^{[20]}$ &MPNN & /  &Spatial &PP& ICLR'19  & \url{https://github.com/klicperajo/dimenet} \\
&SphereNet$^{[21]}$ &MPN & /&Spatial&PP&ICLR'22  &\url{https://github.com/divelab/DIG}\\
&ConfVAE$^{[22]}$ &GNN  &/ &Spatial&MG&ICML'21  &\url{ https://github.com/MinkaiXu/ ConfVAE-ICML21}\\
&GRAPHMVP$^{[23]}$ &GNN &CL &Spatial&PP &ICLR'22 &\url{https://github.com/chao1224/GraphMVP}\\
 &3D-Informax$^{[24]}$ &MPNN &CL & Spatial &PP&ICML'22&\url{https://github.com/HannesStark/3DInfomax}\\
 &UnifiedPML$^{[25]}$ &GN Blocks & RT &Spatial&PP &KDD'22&\url{https://github.com/teslacool/UnifiedMolPretrain}\\
&GeomGCL$^{[26]}$ &MPNN &CL &Spatial&PP &AAAI'22  & /\\
\midrule
\multirow{2}{*}{KG} &KGNN$^{[27]}$ &GNN & / &Drug-drug interaction&DDI &IJCAI'20  &\url{https://github.com/xzenglab/KGNN} \\
&KCL$^{[28]}$ &MPNN &CL &Chemical element &DDI &AAAI'22  &\url{https://github.com/ZJU-Fangyin/KCL}\\
\midrule
\multicolumn{8}{l}{\makecell[l]{ \footnotesize $^{[1]}$\cite{gilmer2017neural}; $^{[2]}$\cite{Hu2020Strategies};$^{[3]}$\cite{sun2020infograph};$^{[4]}$ \cite{brockschmidt2020gnn}; $^{[5]}$\cite{rong2020self};$^{[6]}$\cite{you2020graph};$^{[7]}$\cite{sun2021mocl};\\
\footnotesize $^{[8]}$\cite{zhang2021motif};$^{[9]}$\cite{yang2021deep};$^{[10]}$\cite{liu2022GREA}; $^{[11]}$\cite{yang2022learning};$^{[12]}$\cite{jin2018junction};$^{[13]}$\cite{jin2018learning};$^{[14]}$\cite{shi2020graphaf};\\\footnotesize$^{[15]}$\cite{jin2020multi};$^{[16]}$\cite{jin2020hierarchical};$^{[17]}$\cite{maziarz2022learning};;$^{[18]}$\cite{jin2017predicting};$^{[19]}$\cite{wang2022chemicalreactionaware};$^{[20]}$\cite{klicpera2019directional};$^{[21]}$\cite{liu2022spherical};\\\footnotesize 
$^{[22]}$\cite{xu2021end};$^{[23]}$\cite{liu2022pretraining};$^{[24]}$\cite{stark20223d};$^{[25]}$\cite{zhu2022unified};$^{[26]}$\cite{li2022geomgcl};$^{[27]}$\cite{lin2020kgnn};$^{[28]}$\cite{fang2022molecular};
 }}\\
\bottomrule
\end{tabular}}}
\caption{Representative graph-based MRL algorithms with open-source code. CL and RT stand for pre-training by two self-supervised learning methods, \underline{c}ontrastive \underline{l}earning and \underline{r}econstruction \underline{t}asks respectively. PP, MG, RP and DDI stand for \underline{p}roperty \underline{p}rediction, \underline{m}olecule \underline{g}eneration, \underline{r}eaction \underline{p}rediction and \underline{d}rug-\underline{d}rug \underline{i}nteraction respectively. ``/'' indicates not applicable. } 
\label{tab:main_algorithm}
\end{center}

\end{table*}

\vspace{-0.1in}
\section{Methodologies of MRL}
In this section, we summarize MRL methods into three categories based on the types of input molecules: 2D-based, 3D-based, and knowledge graph-based MRL methods. We introduce the encoding method for each category and point out recent representative methodologies (summarized in Table \ref{tab:main_algorithm}). 

\subsection{2D-based MRL Methods}
2D molecular graphs are the most widely used inputs for graph-based MRL. Here, we introduce the general graph neural networks to learn molecular representations with 2D graphs. Then, following several chemical substructure definition clarifications, we will chart the path from the general representation learning methods to the representative methods incorporating molecular structures and chemistry-related domain knowledge. 

\textbf{Encoding methods.} Formally, each molecule generally is considered as an undirected graph $G=(\mathcal{V}, \mathcal{E}, X)$ with node features $x_v \in X$ for $v\in \mathcal{V}$ and edge features $e_{uv} \in E$ for $(u,v) \in \mathcal{E}$ ~\cite{brockschmidt2020gnn}. Here, nodes represent atoms and edges represent bonds. Generally, graph-based learning methods can fit into Message Passing Neural Networks (MPNN)~\cite{gilmer2017neural} scheme. Therefore, we take MPNN as an example to illustrate the learning process. The forward pass consists of three operations: message passing, node update, and readout. During the message passing phase, node features are updated iteratively according to their neighbors in the graph for $T$ times. By initializing the embedding of node $v$ as $h_v^{0} = x_v$, node hidden states $\textbf{$h$}^{t+1}_v$  at step $t+1$ are obtained based on messages $m_v^{t+1}$, which are represented as: $\textbf{$m$}_v^{t+1} = \sum_{u\in \mathcal{N}(v)} {M_t(h_v^{t},\ h_u^{t},\ e_{uv})}$ and $\textbf{$h$}^{t+1}_v = U_t(h_v^{t},\ m_v^{t+1} ),$
where $M_t$ denotes the message function, $U_t$ is the node update function, and $\mathcal{N}(v)$ is the set of node $v$'s neighbors in the graph. After updating the node features $T$ times, the readout function $R$ computes the whole graph embedding vector by $\hat{y} = R({h_v^T\ |\ v\in \mathcal{V}}).$
Note that $R$ is invariant to the order of nodes so that the framework can be invariant to graph isomorphism. $\hat{y}$ is the representation for the molecule and can be passed to downstream tasks. All functions $M_t$, $U_t$, and $R$ are neural networks with learnable weights to update during the training process. 

Besides MPNN, different variants of graph neural networks like GCN~\cite{kipf2016semi}, GIN~\cite{xu2018powerful}, GAT~\cite{velivckovic2017graph}, GGNN~\cite{li2016gated}, GraphSage~\cite{hamilton2017inductive}, HetGNN~\cite{zhang2019heterogeneous} and feature-wise GNN-Film~\cite{brockschmidt2020gnn} can also be used directly to learn molecular representations. These methods are widely utilized as the base encoder for molecular representation learning in various downstream tasks, such as reaction prediction~\cite{coley2019graph}, property prediction~\cite{brockschmidt2020gnn} and drug discovery~\cite{jin2020multi}. Hu et al.~\cite{Hu2020Strategies} conduct a comparative study on GNNs in property prediction and find that GIN usually achieves the best results. While these models are powerful in learning graph structures, chemical traits, and other chemical domain knowledge are largely neglected. 

\textbf{Chemical Substructures.} Molecular graphs have substructures that are relevant to certain molecular properties or represent  molecular generation constraints. These substructures are not just subgraphs, and convey special  domain knowledge. They are clarified and distinguished next. \emph{Chemical substructures}~\cite{jin2018junction} could be clusters of atoms, e.g., rings. \emph{Motifs}~\cite{rong2020self,sun2021mocl} are recurrent sub-graphs among the input graphs. The \emph{functional group} is an important component of motifs and encodes rich domain knowledge of molecules. This type of motif can be detected by RDKit~\cite{rdkit}. A \emph{Rationale}~\cite{jin2020multi,liu2022GREA} is a sub-graph that has a particular molecular property. \emph{Scaffolds}~\cite{maziarz2022learning} are predefined chemical sub-graphs. Structures sharing the same scaffold can always be considered to be generated following the same synthetic pathway. 

\textbf{Learning Strategies.} 
The weights in MRL encoder can be trained in an end-to-end fashion by attaching the encoder with a downstream task, e.g., the representation after encoding is sent to make property prediction by a fully connected layer. The training thus takes a supervised  manner. To enhance the training process,  molecular substructure-related or chemical domain knowledge can be utilized. JT-VAE~\cite{jin2018junction}, RationaleRL~\cite{jin2020multi}, and HierVAE~\cite{jin2020hierarchical} take advantages of chemical substructures, rationales, and motifs respectively for the molecular generation. Yang et. al~\cite{yang2021deep} proposed PhysChem. This method is composed of a physicist network to learn the molecular conformation and a chemist network to learn the molecular properties. It shows good performance on property prediction benchmarks by fusing both chemical and physics information. Wang et. al~\cite{wang2021property} involved task information and proposed a property-aware embedding method for molecular property prediction. 

Although the learned representation may perform well for the specific downstream task, it is not generally usable for other tasks. In addition, supervised training requires a sufficient set of annotated training samples, which are often difficult to acquire. Recent research has seen a foray into self-supervised learning strategies~\cite{jin2020self} that propose reconstruction tasks for pre-training. PreGNN~\cite{Hu2020Strategies} uses two self-supervised strategies (context prediction and node/edge attribute masking) to pre-train GNN. GROVER~\cite{rong2020self} involves molecular-specific self-supervised methods: contextual property prediction and graph-level motif prediction. Zhang et. al~\cite{zhang2021motif} also designed a motif-based graph self-supervised strategy, which predicted the motif's topology and label during the motif tree generation process.

Contrastive learning is another common self-supervised learning technique, where augmented graphs are biased to keep close to the anchor graph (positive pair) and away from other graphs (negative pairs). It can help graph encoder models to produce graph representations with better generalizability, transferability, and robustness. The general graph augmentation methods (node dropping, edge perturbation, attribute masking and subgraph) were proposed by You et al.~\cite{you2020graph}, which could be applied to molecular datasets as well. InfoGraph~\cite{sun2020infograph} trains the model by maximizing the mutual information between the representations of the entire graph and substructures of different granularity. Unlike general contrastive learning strategies, the following models incorporate chemical domain knowledge. MoCL~\cite{sun2021mocl} has two proposed molecular graph augmentation methods: one is replacing a valid substructure with a similar physical or a chemical property-related substructure. The other one is changing a few general carbon atoms. Wang et. al~\cite{wang2022chemicalreactionaware} were inspired by the relation of equivalence between reactants and products in a chemical reaction. They proposed MolR to preserve the equivalence relation in the embedding space. It forces the sum of reactant embeddings and the sum of product embeddings to be equal. The reactant and the product from different reactions could be a negative pair for this contrastive learning. 

\subsection{3D-based MRL Methods}
 Molecular spatial information, especially geometric information, attracts wide attention and has been increasingly involved in the MRL in recent years, especially when the model needs to learn the physical and chemical properties of molecules associated with the 3D positions of atoms. 

 \textbf{Encoding methods.} 3D molecule is a dynamic structure since atoms are in continual motion in  3D space. The local minima on the potential energy surface are called conformer, or conformation. In nature, a molecule contains multiple low-energy conformers exhibiting different chemical properties. Therefore, the 3D molecular graph implicitly encodes spatial positions of atoms to learn better representation. In general, the 3D conformer of a molecule can be denoted as $G^{3D}=(\mathcal{V}, \mathcal{C})$  where $\mathcal{V}$ is the node set, and $\mathcal{C}$ is the 3D-coordinate matrix. Various encoding methods are explored to enhance GNNs with a proper 3D-coordinate matrix. GeomCL~\cite{li2022geomgcl}  and GRAPHMVP~\cite{liu2022pretraining} utilize conformers generated from RDkit through the stochastic optimization algorithm using the Merck Molecular Force Field (MMFF) to encode spatial information such as angles, inter-atomic distances, torsion, etc. SphereNet~\cite{liu2022spherical} employs sphere coordinates and designs a spherical message passing as a powerful scheme for 3D molecular learning. Follow this work, ComENet~\cite{wang2022comenet} is a message-passing paradigm on 3D molecular graphs with better completeness (bijectivity) by leveraging rotation angles.  To circumvent the challenge that true 3D coordinates of molecules are difficult to calculate and sometimes non-deterministic, DimeNet~\cite{klicpera2019directional}, GemNet~\cite{klicpera2021gemnet} and Directional MPNN~\cite{klicpera2021directional} generate synthetic coordinates in molecules through computing molecular distance bound and corresponding angles between atom pairs, where directional message passing networks are applied to learn the enhanced representation.   

\textbf{Learning Strategies.} In addition to supervised learning, recent MRL studies propose self-supervised learning techniques by using both 2D and 3D molecular graphs. UnifiedPML~\cite{zhu2022unified} uses three pre-training reconstruction tasks: reconstruction of masked atom and coordinates, 3D conformation generation based on 2D graph, and 2D graph generation based on 3D conformation. For contrastive learning, molecular 2D and 3D graph representations are naturally two augmented views of molecules. Using this characteristic, GeomGCL~\cite{li2022geomgcl}, GraphMVP~\cite{liu2022pretraining}, and 3D-Informax~\cite{stark20223d} were proposed to train the molecular representations by keeping the consistency between 2D and 3D graph information. GeomGCL utilizes 2D geometric information, while GraphMVP and 3D-Informax use 2D topological information. Different from the other two methods, 3D-Informax utilizes multiple 3D conformers instead of one.

\subsection{Knowledge Graph-based MRL Methods}
The knowledge graph-based methods are proposed to involve molecular-structure-invariant but rich external knowledge in the model. KCL~\cite{fang2022molecular} is a molecular augmentation method for contrastive learning with an external knowledge graph, which is formed by triples in the form of (chemical element, relation, attribute), such as (Gas, isStateOf, Cl). Then a new node ``Gas'' connecting with ``Cl'' atom will be generated to the original 2D molecular graph. After the augmentation, the model will be trained by maximizing the agreement between two views of molecular graphs with a contrastive loss function. In contrast to KCL and the above MRL methods, which take atoms as nodes and bonds as edges forming a graph, KGNN~\cite{lin2020kgnn} and MDNN~\cite{lyumdnn} explore the knowledge graph consisting of molecules as nodes and connection relationship between molecules as edges. In this case, molecular representations are learned from the knowledge graph structures instead of molecular structures.

\begin{table}[t]
\begin{center}
\scale[0.73]{\begin{tabular}{l|l|c|c|c|c}
\toprule
Methods  & Reference & \multicolumn{3}{c}{Evaluation Metrics} \\ \midrule\midrule
\multicolumn{2}{l|}{\textbf{Property Prediction}} & MAE &RMSE & AUC &ACC\\\midrule
MPNN &\cite{gilmer2017neural} & \checkmark &&\\
DimeNet &\cite{klicpera2019directional} &\checkmark &&\\
Pre-GNN &\cite{Hu2020Strategies} & & &\checkmark &\\
InfoGraph &\cite{sun2020infograph} &\checkmark & & &\checkmark\\
GNN-FiLM &\cite{brockschmidt2020gnn} &\checkmark &&\\
GROVER &\cite{rong2020self} &\checkmark &\checkmark &\checkmark &\\
GraphCL &\cite{you2020graph} & & & &\checkmark\\
MoCL &\cite{sun2021mocl} & & &\checkmark &\\
MGSSL &\cite{zhang2021motif} & & &\checkmark &\\
PhysChem &\cite{yang2021deep} &\checkmark &\checkmark & & \\
KCL &\cite{fang2022molecular} & &\checkmark &\checkmark &\\
GeomGCL &\cite{li2022geomgcl} & &\checkmark &\checkmark &\\
GRAPHMVP &\cite{liu2022pretraining} & & &\checkmark & \\
3D-Informax &\cite{stark20223d} &\checkmark & & &\\
UnifiedPML* &\cite{zhu2022unified} &\checkmark &\checkmark &\checkmark &\\
GREA &\cite{liu2022GREA} & &\checkmark &\checkmark &\\
MoleOOD &\cite{yang2022learning} & & &\checkmark &\\
\midrule\midrule
\multicolumn{2}{l|}{\textbf{Molecular Generation}} & Validity & Diversity & \multicolumn{2}{c}{Others}\\\midrule
JT-VAE &\cite{jin2018junction} &\checkmark & & \multicolumn{2}{c}{Reconstruction}\\
GraphAF &\cite{shi2020graphaf} &\checkmark &\checkmark & \multicolumn{2}{c}{Reconstruction}\\
RationaleRL &\cite{jin2020multi} & &\checkmark & \multicolumn{2}{c}{Success}\\
HierVAE &\cite{jin2020hierarchical} &\checkmark &\checkmark & \multicolumn{2}{c}{Reconstruction}\\
MoLeR* &\cite{maziarz2022learning} &\checkmark &\checkmark &\multicolumn{2}{c}{FCD}\\
ConfVAE &\cite{xu2021end}  & & &\multicolumn{2}{c}{COV\&MAT}\\
UnifiedPML*  &\cite{zhu2022unified} & & &\multicolumn{2}{c}{COV\&MAT}\\
\midrule\midrule
\multicolumn{2}{l|}{\textbf{Reaction Prediction}} \\\midrule
WLDN++* &\cite{coley2019graph} & \multicolumn{4}{c}{Coverage \& Accuracy}\\
MolR* &\cite{wang2022chemicalreactionaware} & \multicolumn{4}{c}{MRR \& Hits}\\
\midrule\midrule
\multicolumn{2}{l|}{\textbf{Drug-drug Interactions}} & AUC &P-AUC & ACC &F1\\\midrule
AttSemiGAE &\cite{ma2018drug} & \checkmark & \checkmark & &\\
KGNN &\cite{lin2020kgnn} & \checkmark & \checkmark &\checkmark &\checkmark\\
MDNN* &\cite{lyumdnn}  & \checkmark & \checkmark &\checkmark &\checkmark\\
\bottomrule
\end{tabular}}
\caption{Applications for MRL with representative methods and evaluation metrics for each method. The methods marked with ``*'' denote SOTA for each application. AUC indicates ROC-AUC.}
\label{tab:application}
\end{center} 
\end{table}

\begin{table*}[t]
\begin{center}
\scale[0.77]{{\begin{tabular}{l|l|l|l|l|l|l}
\toprule
\textbf{Dataset} & \textbf{Category} & \textbf{\#Train} &\textbf{\#Dev} &\textbf{\#Test} & \textbf{Reference} &\textbf{Data Link} \\ 
\midrule
ZINC15 & Structure Pretraining &/ &/ &/ & \cite{sterling2015zinc} & \url{https://zinc15.docking.org} \\
PubChem & Structure Pretraining &/ &/ &/ & \cite{kim2019pubchem} &\url{https://pubchem.ncbi.nlm.nih.gov} \\
ChEMBL & Structure Pretraining &/ &/ &/ &\cite{gaulton2017chembl} & \url{https://www.ebi.ac.uk/chembl/} \\
QM9 & Property prediction &107,108 &13,388 &13,388 & \cite{wu2018moleculenet}&\url{https://moleculenet.org/datasets-1} \\
ESOL & Property prediction &902 &112 &112 & \cite{wu2018moleculenet}&\url{https://moleculenet.org/datasets-1}  \\
FreeSolv & Property prediction &513 &64 &64 & \cite{wu2018moleculenet}&\url{https://moleculenet.org/datasets-1}  \\
Lipophilicity & Property prediction &3,360 &420 &420 & \cite{wu2018moleculenet}&\url{https://moleculenet.org/datasets-1}  \\
MUV & Property prediction &74,470 &9,308 &9,308 & \cite{wu2018moleculenet}&\url{https://moleculenet.org/datasets-1}  \\
HIV & Property prediction &32,901 &4,112 &4,112 & \cite{wu2018moleculenet}&\url{https://moleculenet.org/datasets-1}  \\
PDBbind & Property prediction &9,526 &1,190 &1,190 & \cite{wu2018moleculenet}&\url{https://moleculenet.org/datasets-1}  \\
BACE & Property prediction &1,210 &151 &151 & \cite{wu2018moleculenet}&\url{https://moleculenet.org/datasets-1}  \\
BBBP & Property prediction &1,631 &203 &203 & \cite{wu2018moleculenet}&\url{https://moleculenet.org/datasets-1}  \\
Tox21 & Property prediction &6,264 &783 &783 & \cite{wu2018moleculenet}&\url{https://moleculenet.org/datasets-1}  \\
ToxCast & Property prediction &6,860 &857 &857 & \cite{wu2018moleculenet}&\url{https://moleculenet.org/datasets-1}  \\
SIDER & Property prediction &1,141 &142 &142 & \cite{wu2018moleculenet}&\url{https://moleculenet.org/datasets-1}  \\
ClinTox & Property prediction &1,182 &147 &147 & \cite{wu2018moleculenet}&\url{https://moleculenet.org/datasets-1}  \\
USPTO\_MIT& Reaction Prediction &400,000 & 40,000 & 40,000 &\cite{jin2017predicting} &\url{https://github.com/wengong-jin/nips17-rexgen} \\
USPTO-15K& Reaction Prediction &10500 & 1500 & 3000 &\cite{coley2017prediction} &\url{https://github.com/connorcoley/ochem_predict_nn} \\
USPTO-full & Reaction Prediction &760,000 &95,000 &95,000  &\cite{lowe2012extraction} &\url{https://github.com/dan2097/patent-reaction-extraction}  \\
ZINC-250k & Molecular Generation &200,000 &25,000  &25,000 & \cite{kusner2017grammar} &\url{https://github.com/mkusner/grammarVAE}  \\
DrugBank &Drug-drug interaction &489,910 &61,238 &61,238 &\cite{lin2020kgnn} &\url{https://github.com/xzenglab/KGNN} \\
KEGG-drug &Drug-drug interaction &45,586 &5,698 &5,698 &\cite{lin2020kgnn} &\url{https://github.com/xzenglab/KGNN/tree/master} \\
\bottomrule
\end{tabular}}}
\caption{Datasets used in  MRL study.}
\label{tab:dataset}
\end{center}

\end{table*}

\section{Application Tasks of MRL}

In this section, we discuss four real applications of MRL, and present the details of representative work in Table~\ref{tab:application}.

\subsection{Property Prediction}
Molecular property prediction plays a fundamental role in drug discovery to identify potential drug candidates with target properties. Generally, this task consists of two phases: a molecular encoder to generate a fixed-length molecular representation and a predictor. A predictor is utilized to predict whether the molecule has the target property or predict the reaction of molecules to the target property based on learned molecular representation. Property prediction results can reflect the quality of learned molecular representation directly. General graph learning papers~\cite{Hu2020Strategies,gilmer2017neural,brockschmidt2020gnn,you2020graph} 
thus take property prediction tasks to evaluate the performance of their algorithms. Compared with these general methods, the pre-training methods (e.g., GROVER, MoCL, and MGSSL) involving molecular substructure-related designs are more appropriate 
and can mostly achieve better prediction performance. Some recent work proposes supervised learning models specifically for property prediction. GREA~\cite{liu2022GREA} defines the complementary subgraph of rationale in each molecular graph as an environmental subgraph of that molecule, which has no relationship with the target property. 
Yang et al.~\cite{yang2022learning} proposed MoleOOD to guide the encoder to focus on the environment-invariant substructures to improve its generalization ability. Among the above methods, UnifiedPML~\cite{zhu2022unified} achieves SOTA performance by leveraging the unified view of both 2D and 3D molecular graphs. Besides, the insufficient available molecular dataset is a common problem existing in chemistry. Guo et al.~\cite{guo2021few} and Wang et al.~\cite{wang2021property} proposed meta-learning methods to deal with this low-data problem on property prediction. 

The property prediction task is mainly conducted as classification and regression. The classification task is to predict a discrete class label, for which the Area under the ROC curve (ROC-AUC) and accuracy (ACC) are selected evaluation metrics~\cite{sun2020infograph,rong2020self}. The regression task is to predict a continuous quantity for each molecule, where the mean absolute error (MAE) and Root Mean Square Error (RMSE) are commonly used evaluation metrics to provide the estimate of the accuracy for the target~\cite{rong2020self,zhu2022unified}.

\subsection{Molecular Generation}
The key challenge of drug discovery is to find target molecules with the target properties, which heavily relies on domain experts. The molecular generation is to automate this process. Two steps are necessary to complete this task: one is designing an encoder to represent molecules in a continuous manner, which is beneficial to optimize and predict property; the other is proposing a decoder to map the optimized space to a molecular graph with the optimized property. Atom-by-atom generations may generate atypical chemical substructures such as partial rings. To avoid invalid states~\cite{jin2018junction}, most studies generate graphs fragment by fragment instead of node by node. JT-VAE~\cite{jin2018junction} and VJTNN~\cite{jin2018learning} decompose the molecular graph into the junction tree first, based on substructures in the graph. Then they encode the junction tree using a neural network. Next, they reconstruct the junction tree and assemble nodes in the tree back to the original molecular graph. These methods are highly complex and frequently fail for substructures involving more than 10 atoms. Different from the above substructures, motifs have much larger and more flexible substructures. Therefore, to deal with the problems existing in the previous methods, Jin et al.~\cite{jin2020hierarchical} proposed HierVAE, which relies on motifs instead of substructures and generates molecular graphs hierarchically based on motifs. Following this direction, MoLeR~\cite{maziarz2022learning} was proposed to generate molecules by combining atom-by-atom and motif-by-motif generation. What's more, it supports the scaffolds as an initial seed for molecular generation, which outperforms previous methods on scaffold-constrained tasks. Besides substructures and motifs, RationaleRL~\cite{jin2020multi} utilizes Monte Carlo Tree Search to extract the sub-graphs from the molecules and construct rationale vocabulary for each property with their predictive models. It composes molecules from the sampled rationales to preserve the properties of interest with a variational auto-encoder. In contrast to the above encoder-decoder framework, GraphAF~\cite{shi2020graphaf} is a flow-based auto-regressive model to generate the molecular graph in a sequential process, which allows it to leverage chemical valency constraints in each generation step and also enjoys efficient parallel computation in the training process. ConfVAE~\cite{xu2021end} and UnifiedPML~\cite{zhu2022unified} were proposed for molecular 3D conformation generation tasks. 

For molecular generation tasks, validity is the percentage of the chemically valid generated molecules~\cite{jin2018junction}. Diversity measures the diversity of the generated positive compounds by computing their distance in chemical space. Reconstruction accuracy is utilized to evaluate how often the model can reconstruct a given molecule from the latent embedding~\cite{jin2020hierarchical}. Frechet ChemNet Distance (FCD) is utilized to measure how much the sampled molecules ensemble the training molecules~\cite{maziarz2022learning}. For property-constraints molecular generation tasks, the model also needs to measure the fraction of generated molecules that match the target property. For molecular conformation generation tasks, coverage score (COV) and matching score (MAT) are usually utilized for evaluation~\cite{xu2021end,zhu2022unified}.

\subsection{Reaction Prediction}
Reaction prediction and retrosynthesis prediction are fundamental problems in organic chemistry. Reaction prediction means using reactants to predict reaction products. The process of retrosynthesis prediction is the opposite of reaction prediction. When taking SMILES as input, the reaction prediction task is analogous to a translation task. When taking molecular graphs as input, there are two steps to do both for reaction prediction and retrosynthesis prediction. Like WLDN~\cite{jin2017predicting} and WLDN++~\cite{coley2019graph}, the model needs to predict the reaction center first and then predict the potential products which is the major product. These tasks will be evaluated by coverage (whether the candidates cover the correct product) and accuracy (whether the model can select the correct product). Different from previous work, MolR~\cite{wang2022chemicalreactionaware} formulates the task of reaction prediction as a ranking problem. All the products in the test set are put in the candidate pool. MolR ranks these candidate products based on the embedding learned from given reactant sets, using mean reciprocal rank (MRR) and hits ratio (Hits) as the evaluation metrics.

\subsection{Drug-drug Interactions}
Detecting drug-drug interaction (DDI) is an important task that can help clinicians make effective decisions and schedule appropriate therapy programs. Accurate DDI can not only help medication recommendations but also effectively identify potential adverse effects, which is critical for patients and society. AttSemiGAE~\cite{ma2018drug} predicts DDI by measuring drug similarity with multiple types of drug features. SafeDrug~\cite{yangsafedrug} designs global and local two modules to fully encode the connectivity and functionality of drug molecules to predict DDI. MoleRec~\cite{yang2023molerec} proposes a molecular substructure-aware representation learning strategy for DDI. Both KGNN~\cite{lin2020kgnn} and MDNN~\cite{lyumdnn} build the drug knowledge graph to improve the accuracy of DDI. DDI prediction is evaluated according to Accuracy (ACC), ROC-AUC (AUC), P-AUC (area under the precision-recall-curve), and F1 scores.

\vspace{-0.1in}
\section{Molecular Datasets and Benchmarks}
We summarize representative molecular representation learning algorithms in Table~\ref{tab:main_algorithm}. To conveniently access the empirical results, each paper is attached with code links if available. Encoding algorithms, pre-training methods, and the utilized domain knowledge are also listed. Here, pre-training methods specify contrastive learning and reconstruction tasks we discussed in Section 3. We also present the representative methods for each application and their corresponding evaluation metrics in Table~\ref{tab:application}. The SOTA for each application is also labeled. In addition, we summarize commonly used datasets for different chemical tasks in Table~\ref{tab:dataset}.  

\vspace{-0.1in}
\section{Future Directions}
Graph-based methods for MRL develop fast. Although MRL has achieved satisfactory results  in various applications, there are still some challenges that remain to be solved. We list several future directions for reference.
\subsection{Graph-based MRL with Spatial Learning}
The 3D geometric information attracts great attention recently in graph-based MRL. There are several ways to encode 3D information. One is an equivariant graph neural network, like SE(3)-transformers~\cite{fuchs2020se}. Another category of methods takes relative 3D information as input, like the directional message passing methods~\cite{klicpera2019directional,klicpera2021gemnet} introduced in Section 3, which include distances between atoms and angles between bonds as features to learn geometric information. SphereNet~\cite{liu2022spherical} leverages spherical message passing to learn 3D molecular representation. Nevertheless, how different geometries contribute to molecular representation learning still lacks rigorous justification. There is no established standard spatial information learning method for now. It should be a promising future research direction for MRL.

\subsection{Graph-based MRL with Explainabitity}
The explainability is always a challenge for MRL. To break down the gap between machine learning and chemical science, a well-designed MRL model to produce competitive prediction or generation results on chemical tasks is important but not the end of MRL research. Which molecular features play the most important part in MRL? How can MRL be helpful on explaining the process of reaction?  How can MRL support the transparent generation of new drugs?  
The answers to these questions will facilitate the discovery and innovation in chemical science and engineering, as well as improving the trustworthiness of machine learning methods. 
AttSemiGAE\cite{ma2018drug}, E2E\cite{gao2018interpretable} and GCNN\cite{henderson2021improving} own initial strategies to improve their model's explainability. However, explainable MRL remains a challenging research problem.  

\subsection{Graph-based MRL with Insufficient Data}
Reliable data collection and annotation are time-consuming and expensive via experiments in the laboratory. As a result, 
data scarcity is a common problem in chemistry, and highly hinders the development of MLR. Self-supervised and meta-learning have been considered promising solutions in recent years. Guo et al.~\cite{guo2021few} and Wang et al.~\cite{wang2021property} proposed meta-learning algorithms to deal with few-shot molecule problems, which appeals to some following work. While only specific application tasks have been investigated, novel MRL algorithms 
should be further developed to deal with insufficient data problems.

\vspace{-0.1in}
\section{Conclusion}
Molecular representation learning builds a strong and vital connection between machine learning and chemical science. In this work, we introduce the problem of graph-based MRL and provide a comprehensive overview of the recent progresses on this research topic. To facilitate reproducible research, we take the first step to summarize and release the representative molecular representation learning benchmarks and commonly used datasets for the research community. This survey paper will be a useful resource for researchers in both chemistry and machine learning to advance the study of MRL and other molecular application tasks.

\section*{Acknowledgements}
This work was supported by National Science Foundation under the NSF Center for Computer Assisted Synthesis (C-CAS), grant number CHE-2202693.

\small
\bibliographystyle{named}
\bibliography{ijcai23}

\end{document}